\documentclass[modern]{aastex61}

\usepackage{graphicx}
\usepackage{amsmath}
\usepackage{amssymb}
\usepackage{enumitem}

\newcommand\mybar{\kern1pt\rule[-\dp\strutbox]{.8pt}{\baselineskip}\kern1pt}

\setlist[itemize]{noitemsep, topsep=0pt, leftmargin=*}

\shorttitle{2005 VL1 and Venera 2}
\shortauthors{Loeb}

\begin{document}

\title{Comment on: ``2005 VL1 is not Venera-2''}

\author{Abraham Loeb}

\affiliation{Astronomy Department, Harvard University, 60 Garden St.,
  Cambridge, MA 02138, USA}

\begin{abstract}
  I show that the small differences between the orbital parameters of
  the dark comet 2005 VL$_1$ and the Venera 2 spacecraft (reported in
  arXiv:2503.07972) are of the
  magnitude expected from gravitational deflection by a close
  encounter of Venera 2 with Venus.
\end{abstract}

\section{Introduction}

\citet{Seligman} identified a population of near-Earth objects (NEOs)
that exhibit statistically-significant non-gravitational accelerations
with no coma, and labeled them ''dark comets". \citet{Loeb} reasoned
that one of these objects, 2005 VL$_1$, was at closest approach to
Earth in late 1965, around the time when the Venera 2 spacecraft was
launched to explore Venus. The observed $H$ magnitude of 2005 VL$_1$
is consistent with a high reflectance from the full surface of Venera
2 including its Solar panels. As known for Venera 2, 2005 VL$_1$
arrived within a short distance from Venus, a highly improbable
coincidence ($\lesssim 1\%$) for the orbital phase of a near-Earth
object that does not target a close approach to Venus. Indeed, 2005
VL$_1$'s orbital parameters are similar to the reported values for
Venera 2. Given the area-to-mass ratio of Venera 2, \citet{Loeb}
showed that 2005 VL$_1$'s non-gravitational acceleration and
negligible transverse acceleration match the values expected from
Solar radiation pressure.

Subsequently, \citet{McDowell} as well as Spada (2025, private
communication) argued that the small differences between the orbital
parameters of 2005 VL$_1$ and Venera 2 imply that they are not the
same object. Here, I show that the small differences in orbital
parameters between these objects could have been caused by
gravitational deflection and unrecorded maneuvers during the flyby
near Venus.

\section{Gravitational Deflection near Venus}

The fractional velocity shift of Venera 2 as a result of its passage at
an impact parameter $b$ and a velocity $v$ relative to Venus is given
by~\citep{Binney},
\begin{equation}
\delta\equiv \left({\delta v_\perp\over v}\right)=\left({2GM\over b
  v^2}\right)~,
\end{equation}
where $G$ is Newton's constant and $M=4.9\times 10^{27}~{\rm g}$ is
the mass of Venus. Adopting $v\sim 30~{\rm km~s^{-1}}$ and $b=24\times
10^3
b_{24}$km~\footnote{\url{https://nssdc.gsfc.nasa.gov/nmc/spacecraft/display.action?id=1965-091A}},
we get $\delta\sim 3\%\times b_{24}^{-1}$. The perihelion distance,
$r_{\rm peri}=a(1-e)$, is $0.718~{\rm au}$ for Venera 2 and
$0.698~{\rm au}$ for 2005 VL$_1$. The small difference between these
values by $\sim 3\%$ is similar in magnitude to the expected value of
$\delta$. An inclination change by $\delta$ in radians corresponds to
1.7$^\circ \times b_{24}^{-1}$.

Radio communication with Venera 2 stopped on February 10, 1966, about
17 days before its closest approach to Venus from which there was no
data return$^1$. If the flyby of Venera 2 near Venus was at a lower
speed or distance than assumed above, then the gravitational
deflection and unrecorded maneuvers of the spacecraft near Venus as
well as any additional orbital perturbations over the past 60 years,
could have resulted in the shifts required to match the final orbital
elements of Venera 2 to those of 2005 VL$_1$.

\bigskip
\bigskip
\bigskip
\bigskip
\section*{Acknowledgements}

This work was supported in part by the {\it Galileo Project} at
Harvard University.
 
\bigskip
\bigskip
\bigskip

\bibliographystyle{aasjournal}
\bibliography{t}
\label{lastpage}
\end{document}